\documentclass[twocolumn,twocolumn]{aastex631}

\newcommand{\HH}    {H$_2$}
\newcommand{\nHH}   {$n({\rm H_2})$}
\newcommand{\CsevO} {C$^{17}$O}
\newcommand{\CeigO} {C$^{18}$O}
\newcommand{\tauCO} {\tau_{\rm C^{18}O}}
\newcommand{\fD}    {$f_{\rm D}$}
\newcommand{\dgc}   {R_{\rm GC}}

\shorttitle{ASHES VI: The core-scale CO-depletion}
\shortauthors{Sabatini G. et al. (2022)}
\graphicspath{{./}{figures/}}

\begin{document}

\title{The ALMA Survey of 70~$\mu$m Dark High-mass Clumps in Early Stages (ASHES). VI. The core-scale CO-depletion.}

\author[0000-0002-6428-9806]{Giovanni~Sabatini}
\affiliation{INAF - Istituto di Radioastronomia - Italian node of the ALMA Regional Centre (It-ARC), Via Gobetti 101, I-40129 Bologna, Italy}

\author[0000-0003-2814-6688]{Stefano~Bovino}
\affiliation{INAF - Istituto di Radioastronomia - Italian node of the ALMA Regional Centre (It-ARC), Via Gobetti 101, I-40129 Bologna, Italy}
\affiliation{Departamento de Astronom\'ia, Facultad Ciencias F\'isicas y Matem\'aticas, Universidad de Concepci\'on, Av. Esteban Iturra s/n Barrio Universitario, Casilla 160, Concepci\'on, Chile}

\author[0000-0002-7125-7685]{Patricio~Sanhueza}
\affiliation{National Astronomical Observatory of Japan, National Institutes of Natural Sciences, 2-21-1 Osawa, Mitaka, Tokyo 181-8588, Japan}
\affiliation{Department of Astronomical Science, The Graduate University for Advanced Studies, SOKENDAI, 2-21-1 Osawa, Mitaka, Tokyo 181-8588, Japan}

\author[0000-0002-6752-6061]{Kaho~Morii}
\affiliation{National Astronomical Observatory of Japan, National Institutes of Natural Sciences, 2-21-1 Osawa, Mitaka, Tokyo 181-8588, Japan}
\affiliation{Department of Astronomy, Graduate School of Science, The University of Tokyo, 7-3-1 Hongo, Bunkyo-ku, Tokyo 113-0033, Japan}

\author[0000-0003-1275-5251]{Shanghuo~Li}
\affiliation{Korea Astronomy and Space Science Institute, 776 Daedeokdae-ro, Yuseong-gu, Daejeon 34055, Republic of Korea}

\author[0000-0002-0528-8125]{Elena~Redaelli}
\affiliation{Max-Planck-Institut f{\"u}r extraterrestrische Physik, Gie{\ss}enbachstra{\ss}e 1, D-85749 Garching bei
M{\"u}nchen, Germany}

\author[0000-0003-2384-6589]{Qizhou~Zhang}
\affiliation{Center for Astrophysics, Harvard \& Smithsonian, 60 Garden Street, Cambridge, MA 02138, USA}

\author[0000-0003-2619-9305]{Xing~Lu}
\affiliation{Shanghai Astronomical Observatory, Chinese Academy of Sciences, 80 Nandan Road, Shanghai 200030, People's Republic of China} 

\author[0000-0002-4707-8409]{Siyi~Feng}
\affiliation{Department of Astronomy, Xiamen University, Zengcuo'an West Road, Xiamen, 361005 China} 

\author[0000-0002-2149-2660]{Daniel Tafoya}
\affiliation{Department of Space, Earth and Environment, Chalmers University of Technology, Onsala Space Observatory, 439~92 Onsala, Sweden}

\author[0000-0003-1604-9127]{Natsuko Izumi}
\affiliation{Academia Sinica Institute of Astronomy and Astrophysics, No.1, Section 4, Roosevelt Road, Taipei 10617, Taiwan, Republic of China}

\author[0000-0003-4521-7492]{Takeshi Sakai}
\affiliation{Graduate School of Informatics and Engineering, The University of Electro-Communications, Chofu, Tokyo 182-8585, Japan.}

\author[0000-0002-8149-8546]{Ken’ichi~Tatematsu}
\affiliation{Department of Astronomical Science, The Graduate University for Advanced Studies, SOKENDAI, 2-21-1 Osawa, Mitaka, Tokyo 181-8588, Japan}
\affiliation{Nobeyama Radio Observatory, National Astronomical Observatory of Japan, National Institutes of Natural Sciences, 462-2 Nobeyama, Minamimaki, Minamisaku,
Nagano 384-1305, Japan}

\author[0000-0002-4173-2852]{David Allingham}
\affiliation{School of Mathematical and Physical Sciences, University of Newcastle, University Drive, Callaghan, NSW 2308, Australia}


\begin{abstract}
Studying the physical and chemical properties of cold and dense molecular clouds is crucial for the understanding of how stars form. Under the typical conditions of infrared dark clouds, CO is removed from the gas phase and trapped on to the surface of dust grains by the so-called depletion process. This suggests that the CO depletion factor (\fD) can be a useful chemical indicator for identifying cold and dense regions (i.e., prestellar cores). We have used the 1.3~mm continuum and \CeigO(2-1) data observed at the resolution of $\sim$5000~au in the ALMA Survey of 70~$\mu$m Dark High-mass Clumps in Early Stages (ASHES) to construct averaged maps of \fD~in twelve clumps to characterise the earliest stages of the high-mass star formation process. The average \fD~determined for 277 of the 294 ASHES cores follows an unexpected increase from the prestellar to the protostellar stage. If we exclude the temperature effect due to the slight variations in the NH$_3$ kinetic temperature among different cores, we explain this result as a dependence primarily on the average gas density, which increases in cores where protostellar conditions prevail. This shows that \fD~determined in high-mass star-forming regions at the core scale is insufficient to distinguish among prestellar and protostellar conditions for the individual cores, and should be complemented by information provided by additional tracers. However, we confirm that the clump-averaged \fD~values correlates with the luminosity-to-mass ratio of each source, which is known to trace the evolution of the star formation process.
\end{abstract}

\section{Introduction}\label{sec1:intro}
Although high-mass stars ($M > 8-10$ M$_\odot$) represent a small fraction compared to less massive counterparts, they play a major role in shaping the physical and chemical properties of the interstellar medium (ISM). The formation of H{\tiny II} regions at the end of the high-mass star formation process may favour conditions for triggering a secondary star formation cycle \citep{Elmegreen98}, involving molecular gas that is richer in complex organic molecules (COMs; \citealt{Herbst-vanDishoeck09}), a large number discovered in the hot molecular cores around  massive young stellar objects (mYSO), e.g., \cite{Kurtz00} and \cite{Cesaroni05}. There is also evidence that the Sun was formed in a cluster that originally hosted high-mass stars (e.g., \citealt{Adams10}). Therefore, studying the details of the formation process of high-mass stars is crucial to understand how the chemical composition of the ISM evolves and how life arises from the organic materials produced during the star formation process. 

In the last few decades, several theoretical scenarios have been proposed to describe the high-mass star formation process \citep[e.g.,][]{Bonnell01, McKee02, Tige17, Kumar20, Padoan20}. These scenarios differ in the initial physical assumptions and predict different formation timescales. The identification and systematic study of the early stages of the high-mass star formation process, before the formation of mYSO(s), is hence crucial for distinguishing between the many existing scenarios \citep[e.g.,][]{Zhang09,Zhang11, Wang14, Sanhueza17, Sanhueza19}.

Infrared dark clouds (IRDCs), originally identified in absorption against the galactic background in the mid-infrared (IR) at 8~$\mu$m \citep[e.g.,][]{Perault96, Egan98}, are so far considered the most likely birthplaces of high-mass stars. These are ubiquitous and extended ($>$10~pc) filamentary structures throughout the Galactic disc, which fragment into clumps and cores, with typical sizes of $\sim$1~pc and $\lesssim$0.1~pc, respectively \citep[e.g.,][]{Carey98, Rathborne06, Simon06, Simon06b, Battersby10, Peretto16, Pokhrel18, Li22}. By combining the IR and radio continuum properties obtained from several galactic plane surveys (e.g., MSX, \citealt{Price01}; MIPSGAL, \citealt{Carey09}; RMS, \citealt{Urquhart09}; ATLASGAL, \citealt{Schuller09}; Hi-GAL, \citealt{Molinari10}; CORNISH, \citealt{Hoare12})\footnote{ATLASGAL: The Atacama Pathfinder EXperiment \citep[APEX,][]{Gusten06} Telescope Large Area Survey of the Galaxy; CORNISH: The Coordinated Radio and Infrared Survey for High-Mass Star Formation; Hi-GAL: Herschel \citep{Pilbratt10} InfraRed Galactic Plane Survey; MIPSGAL: Multiband Imaging Photometer \citep[MIPS,][]{Rieke04} Galactic Plane Survey; MSX: Midcourse Space Experiment Survey of the Galactic Plane; RMS: The Red MSX Source Survey;}, clumps can be classified into evolutionary stages. 

As originally reported by \cite{Saraceno96} for the low-mass regime, the high-mass clumps belonging to different phases also lie in different regions of the $L$-$M$ diagram \citep[see][]{Molinari08}, which compares the circumstellar envelope mass ($M$) and the bolometric luminosity ($L$) for a given clump. The luminosity-to-mass ratio ($L/M$) of the clumps increases from the prestellar- to the more evolved H{\tiny II}-stage as a signature of forming mYSOs, and has therefore been used as an additional diagnostic tool to identify clumps at different evolutionary stages (e.g., \citealt{Molinari08, Elia17, Giannetti17_june, Urquhart18, Urquhart22, Sabatini21}). According to this general scheme, clumps that lack 24~and~70~$\mu$m emission also show a lower $L/M$ ratio and are usually associated with the quiescent/prestellar-stage \citep[e.g.,][] {Zhang04, Chambers09, Sanhueza12, Sanhueza13, Sanhueza19, Guzman15}. However, even under these conditions, it is not possible to completely rule out the presence of star-forming activity in these clumps, which can reveal the presence of cores at different evolutionary stages when observed at high resolution
\citep[e.g.,][]{Feng16b, Sanhueza19, Li19, Li20, Morii21, Tafoya21, Sakai22}. 

Additional chemical constraints have been proposed over time to better characterise the evolutionary picture of the high-mass star formation process. Under the typical physical conditions of dense regions in IRDCs, \nHH $\gtrsim$10$^4$ cm$^{-3}$ and $T_{\rm gas}\lesssim$20 K,  a well-known example of chemical constraint is given by the estimates of the CO depletion (e.g., \citealt{Kramer99, Bergin02, Caselli08, Wiles16, Sabatini19, Feng20}), that has been used in particular to identify the youngest clumps \citep[e.g.,][]{Fontani06, Pillai07, Giannetti14}.

How much of CO is depleting on to the surface of dust grains is usually characterised by the depletion factor (e.g., \citealt{Caselli99, Fontani12, Sabatini19}), defined as the ratio between the expected CO/H$_2$ abundance ($X^\mathrm{E}_{\rm CO }$) and the observed one ($X^\mathrm{O}_{\rm CO }$):
\begin{equation}
f_{\rm D} = \frac{X^\mathrm{E}_{\rm CO }}{X^\mathrm{O}_{\rm CO }} = \frac{X^\mathrm{E}_{\rm CO }\:N({\rm H_2})}{N({\rm CO})},
\end{equation}\label{eq:depletion}

\begin{deluxetable*}{l|cccc|ccccccccc}
\tabletypesize{\footnotesize}
\tablecaption{Summary of the physical and chemical properties of the ASHES sources.\label{tab:sample}}
\tablewidth{-3pt}
\tablehead{
\colhead{Clump-ID} & \colhead{$d_\odot$\tablenotemark{a}} & \colhead{$\dgc$\tablenotemark{a}} & \colhead{Mass\tablenotemark{b}} &\colhead{$R_{\rm eff}$\tablenotemark{c}} & \multicolumn{2}{l}{\underline{~~~rms~(mJy beam$^{-1}$)\tablenotemark{d}~~~}}& \colhead{($V^{\rm C^{18}O}_{\rm lsr}$)\tablenotemark{e}} & \colhead{($\sigma^{\rm C^{18}O}$)\tablenotemark{e}} & \colhead{$\gamma$\tablenotemark{f}} & \colhead{($X^\mathrm{E}_{\rm C^{18}O}$)\tablenotemark{g}}& \colhead{\fD\tablenotemark{h}} \\
\colhead{} & \colhead{(kpc)} & \colhead{(kpc)} & \colhead{(M$_\odot$)} &\colhead{(\arcsec)} & \colhead{1.3~mm} & \colhead{\CeigO~(2-1)} & \colhead{(km s$^{-1}$)} & \colhead{(km s$^{-1}$)} & \colhead{} & \colhead{}& \colhead{}\\
\colhead{(1)} &  \colhead{(2)} & \colhead{(3)}& \colhead{(4)}&\colhead{(5)}   &\colhead{(6)} &\colhead{(7)} & \colhead{(8)} & \colhead{(9)} & \colhead{(10)} & \colhead{(11)}& \colhead{(12)}}
\startdata
G010.991--00.082 & 3.7 & 4.91 & 2230 & 27 & 0.115 & 5.150 & $\:\:\:29.5 \pm 0.3$ & $0.7 \pm 0.3$ & 74 & 5.5$\times 10^{-7}$ & 2.8\\
G014.492--00.139 & 3.9 & 4.79 & 5200 & 23 & 0.168 & 5.200 & $\:\:\:41.2 \pm 0.2$ & $0.8 \pm 0.2$ & 72 & 5.7$\times 10^{-7}$ & 4.6\\
G028.273--00.167 & 5.1 & 4.73 & 1520 & 24 & 0.164 & 5.310 & $\:\:\:80.2 \pm 0.2$ & $0.8 \pm 0.2$ & 71 & 5.9$\times 10^{-7}$ & 3.5\\
G327.116--00.294 & 3.9 & 5.63 &  580 & 20 & 0.089 & 4.150 & $-58.8 \pm 0.2$ & $0.7 \pm 0.2$ & 85 & 4.3$\times 10^{-7}$ & 1.9\\
G331.372--00.116 & 5.4 & 4.56 & 1640 & 24 & 0.083 & 4.270 & $-87.8 \pm 0.1$ & $0.6 \pm 0.1$ & 69 & 6.3$\times 10^{-7}$ & 1.6\\
G332.969--00.029 & 4.4 & 5.03 &  730 & 28 & 0.080 & 4.320 & $-66.5 \pm 0.2$ & $0.7 \pm 0.2$ & 75 & 5.3$\times 10^{-7}$ & 1.2\\
G337.541--00.082 & 4.0 & 5.08 & 1180 & 22 & 0.068 & 3.220 & $-54.6 \pm 0.1$ & $0.6 \pm 0.1$ & 76 & 5.2$\times 10^{-7}$ & 2.1\\
G340.179--00.242 & 4.1 & 4.87 & 1470 & 37 & 0.094 & 5.190 & $-51.8 \pm 0.2$ & $0.8 \pm 0.2$ & 73 & 5.6$\times 10^{-7}$ & 1.1\\
G340.222--00.167 & 4.0 & 4.96 &  760 & 19 & 0.112 & 5.490 & $-51.7 \pm 0.1$ & $0.7 \pm 0.1$ & 74 & 5.4$\times 10^{-7}$ & 1.2\\
G340.232--00.146 & 3.9 & 4.98 &  710 & 25 & 0.139 & 5.440 & $-50.5 \pm 0.1$ & $0.8 \pm 0.1$ & 75 & 5.3$\times 10^{-7}$ & 1.7\\
G341.039--00.114 & 3.6 & 5.23 & 1070 & 27 & 0.070 & 3.340 & $-43.4 \pm 0.1$ & $0.7 \pm 0.1$ & 79 & 4.9$\times 10^{-7}$ & 1.1\\
G343.489--00.416 & 2.9 & 5.75 &  810 & 29 & 0.068 & 3.480 & $-28.6 \pm 0.1$&  $0.5 \pm 0.1$ & 87 & 4.1$\times 10^{-7}$ & 1.9
\enddata
\tablecomments{$^{(a)}$ Taken from \cite{Whitaker17}; $^{(b)}$ Derived from the Millimetre Astronomy Legacy Team 90 GHz (MALT90) Survey \cite{Contreras17}; $^{(c)}$ The clump's effective radius was derived in \cite{Sanhueza19} from Gaussian fitting to the ATLASGAL dust continuum emission at 870~$\mu$m; $^{(d)}$ The rms of dust continuum emission at 1.3~mm are taken from \cite{Sanhueza19}, while those of \CeigO~are computed from the data-cubes presented in Sect.~\ref{sec2:sample}; $^{(e)}$ Median local standard of rest velocities ($V^{\rm C^{18}O}_{\rm lsr}$) and the velocity dispersions ($\sigma^{\rm C^{18}O}$) obtained from the \CeigO~(2-1) employing the Python Spectroscopic Toolkit ({\verb~PySpecKit~}; \citealt{Ginsburg11a, Ginsburg22}; see also Appendix~\ref{app:NC18O}); $^{(f)}$ Gas-to-dust ratio derived using Equation~(\ref{eq:gamma}) (see Sect.~\ref{sub:NHH}); $^{(g)}$ Expected \CeigO/\HH~abundance derived using Equation~(\ref{eq:expected_ab}) (see Sect.~\ref{sub:fD_derivation}); $^{(h)}$ Average \fD~of each clump determined following the procedure discussed in Sect.~\ref{sec4:results};
}                                           
\end{deluxetable*}
\vspace{-32pt}
\noindent
where $N$(\HH) and $N$(CO) are the \HH~and CO column density, respectively. CO-depletion factors of up to a few tens have been derived on clump-scale in various samples of young, high-mass star-forming regions \citep[e.g.,][]{Thomas08, Fontani12, Feng16, Feng20}. 
The estimation of \fD~could be a suitable and convenient way to identify the cold/prestellar gas also at core-scales. However, very few and isolated estimates of \fD~on these scales are found in the literature in high-mass star-forming regions, with extreme values of \fD~up to 100-1000 \citep{Zhang09, Morii21, Rodriguez21}.
In the absence of additional evidence for the high-mass regime, in this study we aim to test whether the CO-depletion factor can be considered a reliable tracer for cores at different evolutionary stages, embedded in high-mass star-forming regions.

This work is structured as follows: in Section~\ref{sec2:sample} we describe the sample and the dataset on which this study is based; in Section~\ref{sec3:analisys} we report on the derivation of the maps of $N$(\HH)~and \CeigO~used to construct the final \fD~maps. In Section~\ref{sec4:results} we discuss the variation in the averaged \fD~obtained for a population of cores at different evolutionary stages; finally, in Section~\ref{sec5:conclusion} we summarise our conclusions.

\section{Sample and data reduction}\label{sec2:sample}
The ALMA\footnote{The Atacama Large Millimeter/submillimeter Array (ALMA; \citealt{Wootten09}).} Survey of 70~$\mu$m Dark High-mass Clumps in Early Stages (ASHES; \citealt{Sanhueza19}) provides an ideal basis for detailed studies of the earliest stages of the high-mass star formation process. In a pilot study \citep{Sanhueza19}, 12 massive 70~$\mu$m dark clumps were mosaicked with ALMA in the dust continuum at $\sim$224 GHz ($\sim$1.2\arcsec~resolution), and used to characterise clump fragmentation (Table~\ref{tab:sample}). We refer to \cite{Sanhueza19} for a detailed description of the source selection criteria. From the dust continuum, a total of 294 cores were detected (excluding those located at the edges of the observed fields -- i.e., $\sim$20-30\% power point -- where flux estimates are more uncertain)\footnote{The complete catalogue is available at \url{https://cdsarc.cds.unistra.fr/viz-bin/cat/J/ApJ/886/102}.}. ASHES was designed to map the molecular emission of a large number of molecules in the ALMA Band-6, including CO, C$^{18}$O, H$_2$CO, CH$_3$OH, SiO, $^{13}$CS, N$_2$D$^+$, DCN, DCO$^+$, and CCD. These tracers are used to characterise cores from a chemical point of view, allowing their classification into different evolutionary stages \citep[see][]{Li20, Morii21,Tafoya21, Sakai22, LiPREP}.

Of the total population of 294 cores, $\sim$71\% of cores (210 cores) are classified as prestellar, lacking of any star formation signatures, while  $\sim$29\% (84 cores) are classified as protostellar candidates, being associated with molecular outflows and/or ``warm core'' line emission (i.e., H$_2$CO and CH$_3$OH lines with high upper energy levels).

Since different chemical conditions were assumed for the identification of the protostellar cores, they were additionally divided into three categories \citep{Sanhueza19, Li20}: (1) cores with molecular outflows (i.e., 24 cores, corresponding to $\sim$8\% of the total population) identified via CO, SiO and/or H$_2$CO lines and in which no ``warm cores'' lines were detected; (2) ``warm cores'' (34 cores, $\sim$12\%) representing an evolutionary phase prior to the hot molecular core phase typically found to be associated with high-mass protostars. This class lacks in molecular outflow emission but shows a detection in one of the ``warm cores'' lines among H$_2$CO \mbox{J = 3$_{2,2}$-2$_{2,1}$} ($E_{\rm u}$/k$_{\rm B}$ = 68.09~K; where k$_{\rm B}$ is the Boltzmann's constant) and \mbox{J = 3$_{2,1}$-2$_{2,0}$} ($E_{\rm u}$/k$_{\rm B}$ = 68.11~K), and the CH$_3$OH \mbox{J$_k$ = 4$_{2,2}$-3$_{1,2}$} ($E_{\rm u}$/k$_{\rm B}$ = 45.46 K); (3) the remaining 26 cores (i.e., 9\%) presumably belong to a more evolved protostellar stage with both molecular outflow and ``warm cores'' line detection.

So far, the ASHES project gives access to the largest population of prestellar cores candidate detected in high-mass star-forming clumps via a mix of dust continuum and line emission data, and reveals that even high-mass 70~$\mu$m dark clumps can harbor a tiny fraction of deeply embedded cores with nascent star formation activity.

\subsection{Observations}\label{sec:observation}
ALMA Band-6 observations were carried out in Cycles 3 and 4 (Project 2015.1.01539.S PI: P. Sanhueza) with the 12m-Array (Main Array, MA; \citealt{Wootten09}) and the 7m-Array (Atacama Compact Array, ACA; \citealt{Iguchi09}). Depending on the observed source, the 12m-array included 36-48 antennas distributed over baselines ranging between 15 and 700~m. The Atacama Compact Array includes 7-10 antennas, with baselines between 8 and 48 m. The average angular scales covered by these configurations range from a resolution $\sim$1.25\arcsec~to a maximum recoverable scale of $\sim$19\arcsec~. These scales correspond to spatial scales of $\sim$(5-70)$\times$10$^3$~au at the average distance of 4~kpc.

At the frequency of the \CeigO(2-1) line, \mbox{$\nu_{\rm 2,1} \sim 219.5$ GHz}, the typical 1$\sigma$ is 5~mJy beam$^{-1}$ (see Table~\ref{tab:sample}), for a channel width of $\sim$0.67~km s$^{-1}$. The data were calibrated with the Common Astronomy Software Applications (CASA) version 4.5.3, 4.6, and 4.7, while the CASA version 5.4 was employed for imaging. The \CeigO~cubes were produced using the automatic masking procedure {\verb~yclean~} \citep{Contreras18}. We refer to \cite{Sanhueza19} for a more detailed description of the dataset.

\section{Analysis and results}\label{sec3:analisys}
Based on Equation~(\ref{eq:depletion}), the derivation of \fD~requires the evaluation of the \HH~and CO column densities. Since the main CO isotopologue (i.e., $^{12}$C$^{16}$O) is almost always optically thick \citep[e.g.,][]{Heyer15}, its intensity is not proportional to $N$(CO). Therefore, a less abundant CO isotopologue (i.e., C$^{18}$O) should be used to obtain a much more accurate estimate of \fD. In this Section, we summarise the procedure and the assumptions we follow to derive \fD.

\subsection{{\rm \HH}~column density maps}\label{sub:NHH}
The beam-averaged \HH~column density is computed in each pixel from the primary beam (PB) corrected ALMA continuum flux density at 1.3 mm, $F_{{\rm 1.3\:mm}}$, as (e.g., \citealt{Schuller09})
\begin{eqnarray}
N({\rm H}_2) = \frac{F_{{\rm 1.3\:mm}}  \gamma}{B_{{\rm 1.3\:mm}}(T_{\rm dust}) \; \Omega_\mathrm{app} \; \kappa_{{\rm 1.3\:mm}} \; \mu_\mathrm{H_2} \; m_\mathrm{H}},
\label{eq:NH2}
\end{eqnarray}

\noindent where $B_{{\rm 1.3\:mm}}(T_{\rm dust})$ is the Planck function at 1.3 mm with a dust temperature $T_{\rm dust}$, $\Omega_\mathrm{app}$ is the beam solid angle\footnote{This is calculated assuming an equivalent radius for a circular beam with the same area as the ALMA beam.}, $\mu_\mathrm{H_2}=2.8$ is the H$_2$ mean molecular weight (\citealt{Kauffmann08}; see their Sect.~A.1), and $m_\mathrm{H}$ is the mass of the hydrogen atom. We adopt a value of $\kappa_{{\rm 1.3\:mm}}=0.9$ cm$^2$ g$^{-1}$, which corresponds to the opacity of thin-icy-mantle dust grains at gas densities of 10$^6$ cm$^{-3}$ \citep{Ossenkopf94}.

In Equation~(\ref{eq:NH2}) we assume $T_{\rm dust}$ equal to the NH$_3$ kinetic temperature, $T_{\rm kin}^{\rm NH_3}$, derived from NH$_3$ (1, 1) and (2, 2) transition lines obtained as part of the CACHMC survey (the Complete ATCA\footnote{The Australia Telescope Compact Array (ATCA; e.g., \citealt{Wilson11}).} Census of High-Mass Clumps; \citealt{AllinghamPREP}) at $\sim$5\arcsec angular resolution. Under typical conditions prevailing in IRDCs, the gas-dust thermal coupling is effective in regions where the gas density exceeds 10$^{4.5}$~cm$^{-3}$ \citep[e.g.,][]{Goldsmith01}. Overall, this density threshold is fulfilled in the entire population of cores identified in ASHES \cite[e.g.,][see also the additional discussion in Sect.~\ref{sub:fD_derivation}]{Sanhueza19}. The methodology to derive the temperature from the NH$_3$ observations is based on \cite{Mangum15} and will be presented in a forthcoming paper describing the survey (\citealt{AllinghamPREP}; see also \citealt{Friesen09, Hogge18} and \citealt{Keown19}). The temperature maps are finally regridded to the same pixel size as the ALMA maps. We mask the native temperature maps where the error is greater than 20\% of the measured $T_{\rm kin}^{\rm NH_3}$, adopting for these pixels the median $T_{\rm kin}^{\rm NH_3}$ temperature of all pixels with emission in the 1.3 mm dust continuum above 3$\sigma$ ($\sigma$ obtained from \citealt{Sanhueza19}; see Table~\ref{tab:sample}). In each source the $T_{\rm kin}^{\rm NH_3}$ ranges from $\sim$7~K to $\sim$50~K, showing on average mild temperature gradients that only in some rare cases reach a few tens Kelvin degrees within the same source. We find an average error of $\sim$12\% $T_{\rm kin}^{\rm NH_3}$, which corresponds to $\sim$2K. 

The gas-to-dust ratio, $\gamma$, is computed following \cite{Giannetti17_oct} with a gradient of $\gamma$ through the Galactic disc.

\begin{equation}\label{eq:gamma}
    {\rm log}_{10}(\gamma) = 0.087\:R_{\rm GC} + 1.44\,,
\end{equation}

\noindent where $R_{\rm GC}$ is the galactocentric distance of each source expressed in kpc (Table~\ref{tab:sample} and \citealt{Whitaker17}). This prescription gives values of the gas-to-dust ratio between 69 and 87 (Table~\ref{tab:sample}), and represents the second modification to the procedure followed by \cite{Sanhueza19} to derive $N$(\HH), where $\gamma$ is taken to be 100. In the worst case scenario, this has produced a modest difference of 30\% in the final $N$(\HH), leaving however unchanged their gradient across the sources. This variation agrees with the intrinsic error of 32\% derived in \cite{Sanhueza17} considering the uncertainties associated with the dust opacity and $\gamma$ in the mass determination of cores, and also reflects the typical error associated with $N$(\HH) considering the uncertainties on $T_\mathrm{dust}$ (e.g., \citealt{Urquhart18, Sanhueza19}). For this reason, we refer to \cite{Sanhueza19}, for the discussion on the distribution of $N$(\HH) in each source and for the visual inspection of the ALMA Band-6 continuum maps.

\subsection{{\rm \CeigO}~column density maps}\label{sub:NCO}
We derive the \CeigO~column density, $N$(\CeigO), from its \mbox{J = 2-1} molecular transition observed with ALMA (see Sect.~\ref{sec:observation}) by following \cite{Kramer91}
\begin{eqnarray}\label{eq:nc18o}
N({\rm C^{18}O}) =  \frac{3h\:C_\tau}{\eta_{\rm c}\:8\pi^3\mu^2}\:f(T_{\rm ex}^{\rm C^{18}O}) \int T_{\rm b}\: d\upsilon,
\end{eqnarray}

\noindent 
with, 
\begin{eqnarray}\label{eq:nc18o_2}
f(T_{\rm ex}^{\rm C^{18}O}) = \frac{\mathcal{Z}}{2}\:\exp\left(\frac{E_{\rm l}}{k_{\rm B} T_{\rm ex}}\right)\left[1-\exp\left( -\frac{h\nu_{\rm 2,1}}{k_{\rm B} T_{\rm ex}}\right)\right]^{-1}\\
\times [J(T_{\rm ex}, \nu_{\rm 2,1}) - J(T_{\rm bg}, \nu_{\rm 2,1})]^{-1}\,, \nonumber
\end{eqnarray}


\noindent
where $h$ is the Planck constant, and $\eta_{\rm c}$ the beam filling factor assumed equal to 1. Also, \mbox{$\mu = 0.112 \times 10^{-18}$ dyn$^{0.5}$ cm$^2$} is the  \CeigO\ dipole moment, $\mathcal{Z} = 0.36\:T_{\rm ex}^{\rm C^{18}O} + 1/3$ is the partition function \citep[e.g.,][]{Herzberg45}, $E_{\rm l}$ is the energy of the lower level of the transition, $T_{\rm ex}^{\rm C^{18}O}$ is gas excitation temperature of \CeigO, $J(T_{\rm ex},\nu)= (h\nu/k_{\rm B})($exp$(h\nu/k_{\rm B} T_{\rm ex})-1)^{-1}$, $T_{\rm bg}=2.7$ K the background temperature, and $T_{\rm b}$ the brightness temperature of the line. The integrated intensity is taken by considering the emission above the 3$\sigma$ threshold in a range of $\pm$5~km s$^{-1}$ around the $V_{\rm lsr}$ reported in Table~\ref{tab:sample}. 

In Equation~(\ref{eq:nc18o}), $f(T_{\rm ex}^{\rm C^{18}O})$ incorporates all the constants and the terms which depend on $T_{\rm ex}^{\rm C^{18}O}$. Also, \mbox{$C_\tau = \tauCO/[1 - {\rm exp}(-\tauCO)]$} is the optical depth correction factor, valid for $\tau\leq2$ with uncertainty of about 15\% (e.g., \citealt{Frerking82, Kramer91}), where $\tauCO$ is the optical depth of the \CeigO(2-1) line derived following the approach discussed in \cite{Sabatini19}, and summarised in Appendix~\ref{app:NC18O}. All the molecular parameters are taken form the Cologne Database for Molecular Spectroscopy (CDMS\footnote{\url{https://cdms.astro.uni-koeln.de/cdms/portal/}}; \citealt{Muller01}). In each source, the 1$\sigma$ rms is computed as the average over five channels -- far from the \CeigO(2-1) line -- of the flux' standard deviations computed in a large region centred at the position of the source. We solve Equation~(\ref{eq:nc18o}) under the assumption of local thermodynamic equilibrium (LTE), i.e., \mbox{$T_{\rm kin}^{\rm NH_3}$ = $T_{\rm dust}$ = $T_{\rm ex}^{\rm C^{18}O}$}. The only exception is G332.969–00.029 that lacks in available NH$_3$ data, and for which we assume $T_{\rm dust}$ = $T_{\rm ex}^{\rm C^{18}O} = 12.6$~K, as the dust temperature reported by \cite{Guzman15}. To avoid possible overestimates of $N$(\CeigO), produced by too low \mbox{$T_{\rm ex}^{\rm C^{18}O} = T_{\rm kin}^{\rm NH_3}$} values, we impose a lower limit  of \mbox{$T_{\rm ex}^{\rm C^{18}O}=$ 10.8 K} that corresponds to the separation between the levels of the \CeigO~(2-1) transition. In the worst-case scenario, this prescription affects less than 6\% of the pixels where $N$(\CeigO) is computed. The opacity-corrected column density map of \CeigO, also corrected for the primary beam effects, are shown in Appendix~\ref{app:NC18O}. Our correction has increased $N$(\CeigO) by up to a factor of about $\sim$1.8, producing $N$(\CeigO) spanning the range \mbox{of $\sim$(0.1-6.4)$~\times~$10$^{16}$~cm$^{-2}$}.

\begin{figure*}
   \centering
   \includegraphics[width=1\hsize]{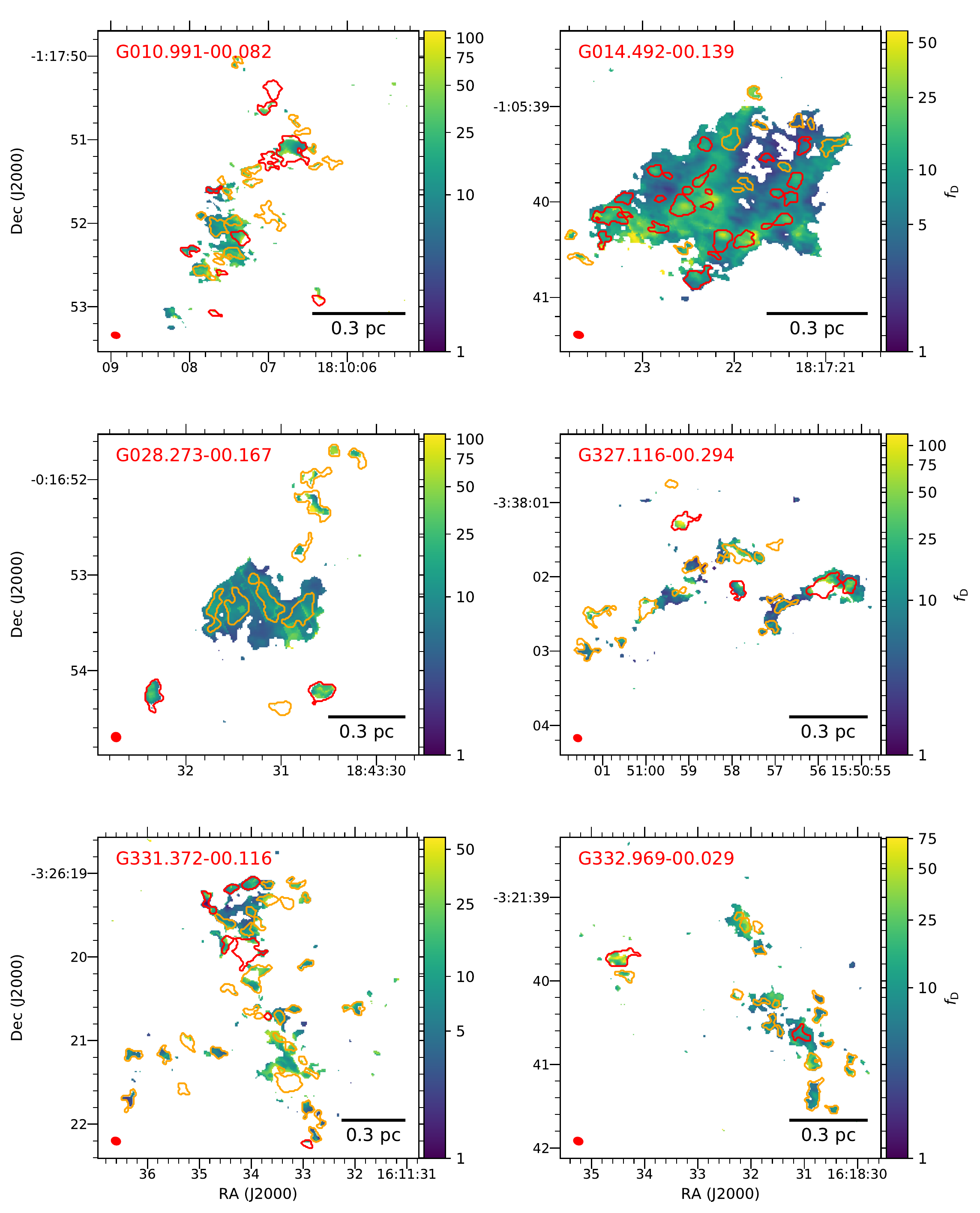}
\caption{ASHES \fD~maps obtained following the procedure explained in Sect.~\ref{sub:fD_derivation}. The cores identified in \cite{Sanhueza19} are shown as orange and red contours for prestellar and protostar cores, respectively, following the classification in Sect.~\ref{sec:observation}. The ALMA synthesized beams are displayed in red at the bottom left in each panel, while the scalebar is shown in the bottom right corners. The color wedge of each panel displays the color scales corresponding to \fD~in the log-scale.}\label{fig:fD_A}%
\end{figure*}

\begin{figure*}
   \centering
   \includegraphics[width=1\hsize]{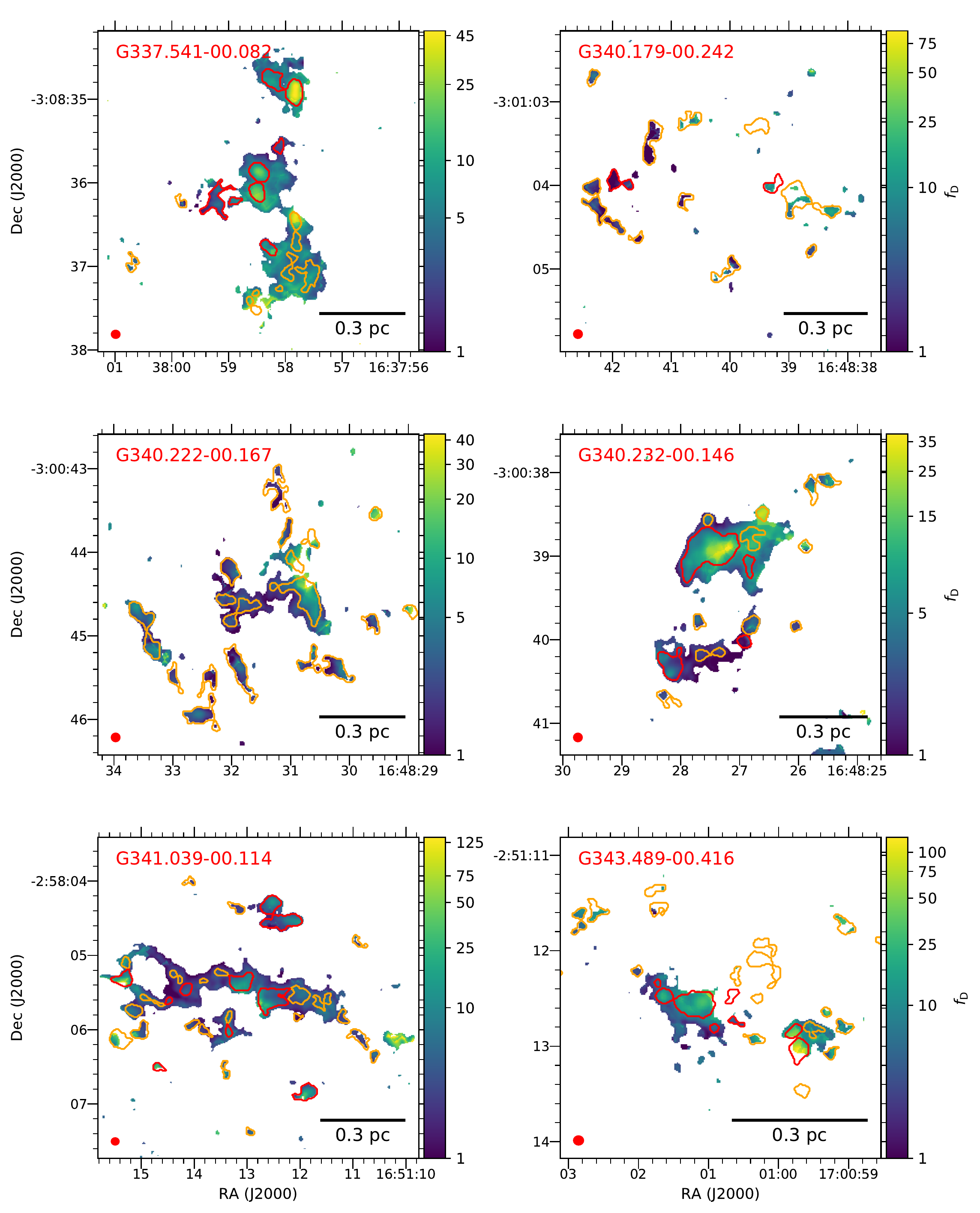}
\caption{Continuation of Fig.~\ref{fig:fD_A}.}\label{fig:fD_B}%
\end{figure*}

\subsection{Core-scale {\rm CO}-depletion maps}\label{sub:fD_derivation}
The final CO-depletion factor maps, shown in Figures~\ref{fig:fD_A} and~\ref{fig:fD_B}, are generated as the ratio between the expected and the observed abundance of CO relative to \HH, following Equation~(\ref{eq:depletion}). For each source, we derive the expected \CeigO/\HH~abundance assuming \citep{Frerking82, Fontani06, Giannetti17_oct}:
	\begin{equation}
		X_\mathrm{C^{18}O}^\mathrm{E} = \frac{9.5 \times 10^{-5} \times 10^{\alpha(\dgc - R_\mathrm{GC,\odot})}}{^{16}\mathrm{O}/^{18}\mathrm{O}} , \label{eq:expected_ab}
	\end{equation}
\noindent
with $\dgc$ expressed in kpc, $R_\mathrm{GC,\odot}=8.34$ kpc \citep{Reid14}, and $\alpha = -0.08\:{\rm dex\:kpc}^{-1}$ describes the C/H abundance \citep{LuckLambert11}, under the assumption that the C/H abundance controls the CO formation\footnote{Note that the C/H abundance is given in dex units, which introduces the term 10$^\alpha$ in Equation~\ref{eq:expected_ab}}. The oxygen isotopic ratio, \mbox{$^{16}\mathrm{O}/^{18}\mathrm{O} = 58.8 \dgc + 37.1$}, is computed according to \cite{WilsonRood94}. We employ the galactocentric distances of the sources reported by \cite{Whitaker17}, according to \cite{Sanhueza19}. We find $X^E_{\rm C^{18}O}$ between $4.1\times 10^{-7}$ and $6.3\times 10^{-7}$ (see Table~\ref{tab:sample}). 

In each \fD~map (Figures~\ref{fig:fD_A} and~\ref{fig:fD_B}) we also report as orange/red regions the cores identified in \citet{Sanhueza19}: orange for the pre-stellar stage and red for the protostellar one (see Sect.~\ref{sec2:sample}).

\begin{figure}
   \centering
   \includegraphics[width=0.95\hsize]{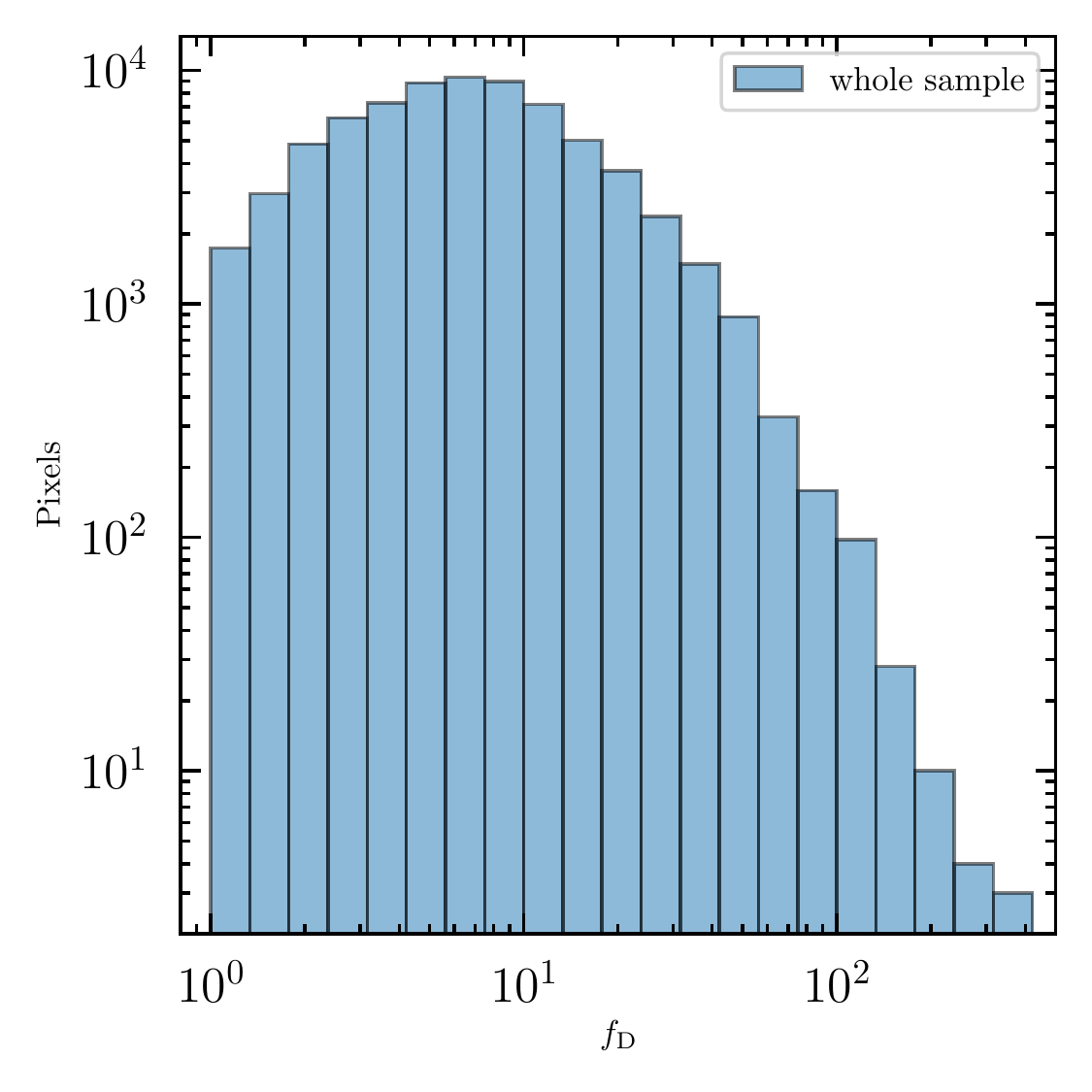}
\caption{Number distribution of \fD~mapped in Figures~\ref{fig:fD_A} and~\ref{fig:fD_B}. Pixel size is 0.2\arcsec, while beam size is 1.2\arcsec.}\label{fig:fDhist_tot}%
\end{figure}
\begin{deluxetable}{lc|c}
\savetablenum{2}
\tablewidth{\columnwidth}
\caption{Averaged \fD~computed for the whole core sample \label{tab:fD_cores}}
\tablehead{
\colhead{Clump-ID}  & \colhead{Core-ID\tablenotemark{a}} & \colhead{$\langle f_{\rm D} \rangle_{\rm core}$}}
\startdata
G010.991-00.082 & ALMA1 & 23.7 \\
G010.991-00.082 & ALMA2 & 92.8 \\
G010.991-00.082 & ALMA3 & 31.9 \\
G010.991-00.082 & ALMA4 & ...  \\
G010.991-00.082 & ALMA5 & 30.9 \\
G010.991-00.082 & ALMA6 & 56.3 \\
G010.991-00.082 & ALMA7 & 11.9 \\
G010.991-00.082 & ALMA8 & 27.6 
\enddata
\tablecomments{$^{(a)}$ The classification of cores follows that defined in \cite{Sanhueza19}. This table is available in its entirety in machine-readable form.}
\end{deluxetable}

The degree of depletion reveals widely different chemical conditions within the individual clumps. It spans regions where CO adsorption is almost irrelevant, with observed abundances of \CeigO~ as expected (i.e., \fD~=~1), up to regions where only less than 1\% of the expected CO is still present in the gas phase (i.e., \fD~$>$~100)\footnote{Note that some of the cores identified in dust continuum are not associated with a value of \fD~(e.g., see G010.991-00.082 and G327.116-00.294 in Fig.~\ref{fig:fD_A}). However around those peculiar regions we find \fD~among the highest over the entire clump. Thus, we expect extremely low abundances of \CeigO\ within those cores (therefore high \fD), below the limit of detection accessible to our observations (Table~\ref{tab:sample}).}. This is clear from Figure~\ref{fig:fDhist_tot} that shows the number distribution of the CO-depletion values over the entire sample. On average, in $\sim$85\% of the area mapped in \fD, more than 50\% of the expected CO has been removed from the gas phase (i.e., \fD$>$2).
This is particularly relevant for the regions within the identified cores (both pre- and proto-stellar).

\begin{figure}
   \centering
   \includegraphics[width=1\hsize]{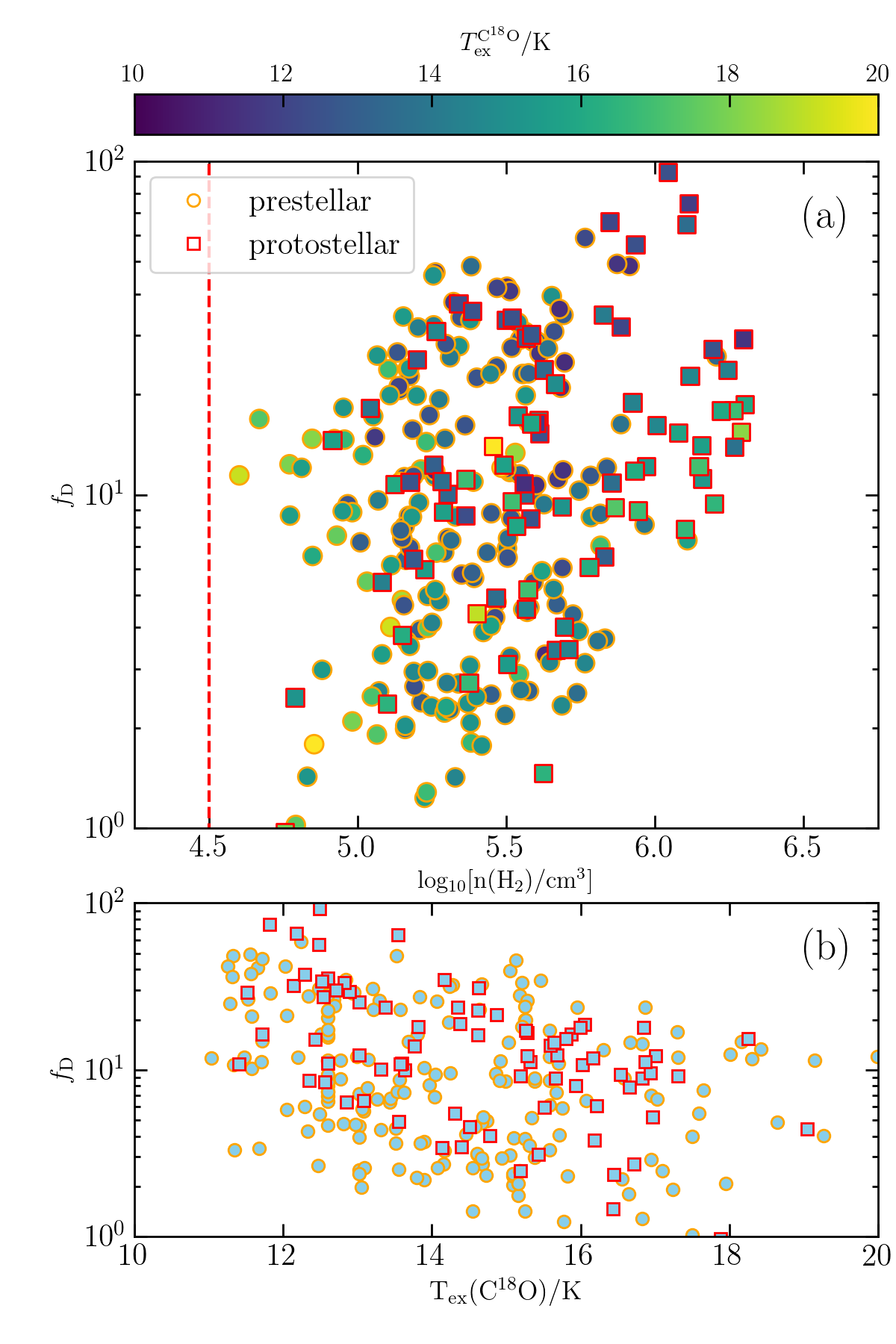}\quad
\caption{Panel (a): Scatter plot of the average properties of the cores identified in ASHES. The orange circles represent the prestellar cores, while the red squares refer to the cores classified as protostellar (Sect.~\ref{sec2:sample}). The red-dashed line represents the density threshold proposed by \cite{Goldsmith01} to ensure the gas-dust thermal coupling. Panel (b): Average \fD~versus $T_{\rm ex}^{\rm C^{18}O}$. The gas excitation temperature of \CeigO~is assumed to be equal to the kinetic temperature of NH$_3$.\label{fig:scatter}}
\end{figure}

In the twelve sources shown in Figures~\ref{fig:fD_A} and~\ref{fig:fD_B}, the behaviour of \fD~follows an unexpected increase as the evolution of the cores progresses, i.e., going from pre- to proto-stellar (see Table~\ref{tab:fD_cores}). As an example, G014.492-00.139 (Fig.~\ref{fig:fD_A}), with a large population of protostellar cores (i.e., 25 protostellar vs 12 pre-stellar cores), shows CO depletion that is comparable (and in some cases higher than) to clumps dominated by pre-stellar cores (e.g., G028.273-00.167 and G340.222-00.167; Fig.~\ref{fig:fD_A} and~\ref{fig:fD_B}, respectively). This can be explained by exploring the physical conditions in the clumps. Due to the absence of protostars and outflows that can heat the gas surrounding the cores, these young clumps reveal mild temperature gradients. Looking at the regions associated with a $>3\sigma$ continuum emission, G028.273-00.167 and G340.222-00.167 show a $\Delta T_{\rm kin}^{\rm NH_3}\sim$1 K, and $\sim$3 K, respectively. If we then neglect the effect of temperature on the desorption process, it is reasonable that the evolution of the averaged \fD\ is mainly density-driven. This is confirmed by the results reported in Fig.~\ref{fig:scatter}(a), which show the depletion factor as a function of both density and temperature. This result also seems independent of the heliocentric distance associated with each ASHES clump, as discussed in Appendix~\ref{app:additioal_plots}. From the same figure we can also see a weak temperature-effect when keeping the density constant, due to the weak temperature gradients in the $T_{\rm kin}^{\rm NH_3}$ maps (see also Fig.~\ref{fig:scatter}b).

\begin{figure}
   \centering
   \includegraphics[width=0.95\hsize]{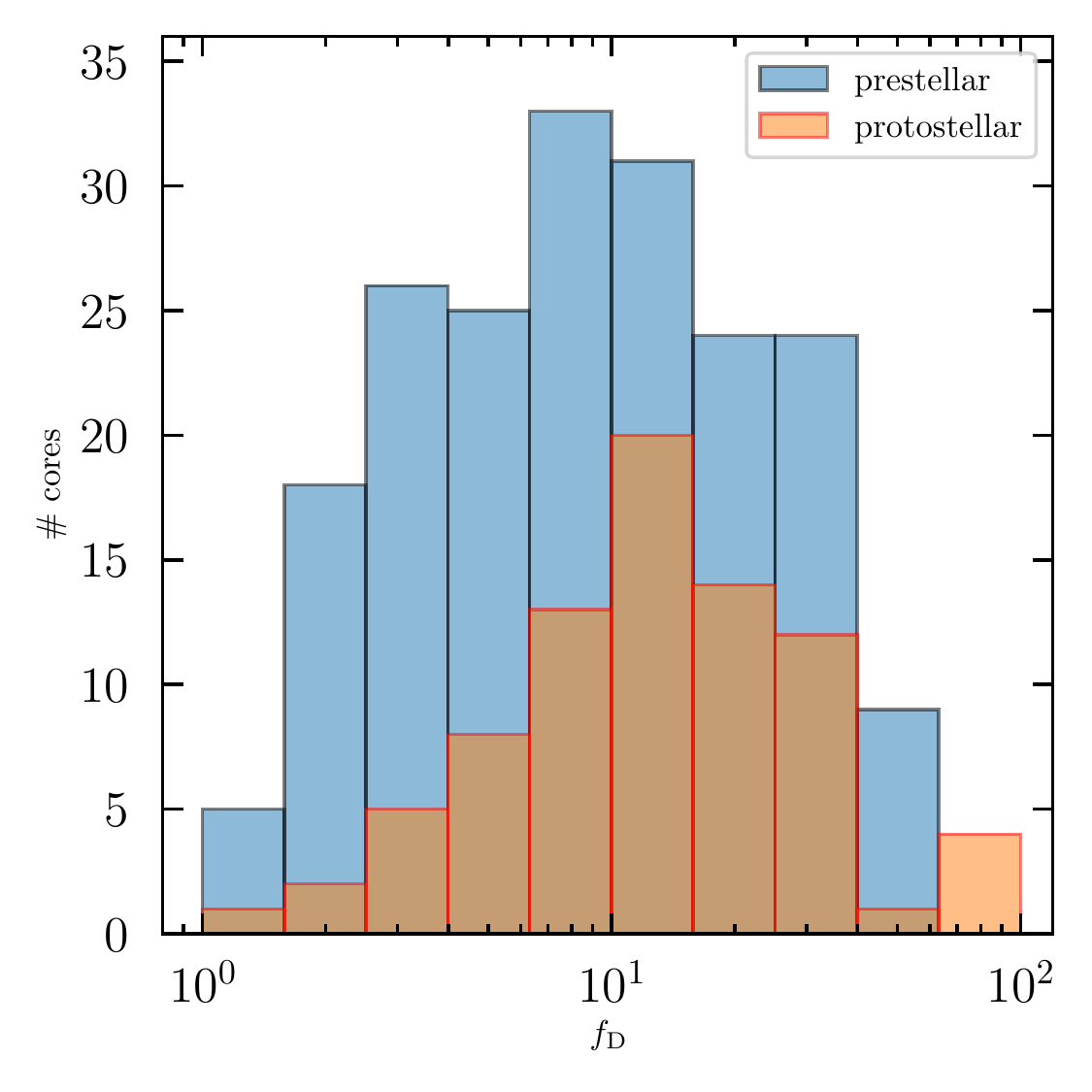}
\caption{Number distributions of averaged \fD~associated to each core identified in ASHES. The blue and orange histograms refer to prestellar and protostellar cores, respectively. A Kolmogorov-Smirnov two-sample test applied to the two distributions yields a $p$-value~$< 2\times 10^{-3}$.}\label{fig:hist}%
\end{figure}

Within an evolutionary picture, clumps dominated by protostellar conditions, are characterized by densities on average larger than the pre-stellar ones (e.g., \citealt{Konig17, Elia21} and \citealt{Urquhart22}). The same evolutionary trend is also seen at cores-scale (\citealt{Sanhueza19}). In addition, all the sources have been selected as 70~$\mu$m dark and it is conceivable that the embedded protostars are at their early stages, i.e., yet to heat the surrounding gas significantly. It is then likely that the high degree of CO depletion is associated with the envelopes of these young protostellar objects, where protostellar activity has not yet led to a significant desorption of the frozen-out CO.

In Fig.~\ref{fig:hist} we show the distributions of \fD~for the entire population of pre- and proto-stellar cores in our sample; blue and orange histograms, respectively. The former show a median \fD~of 8.5, while in the protostellar population we derive a median of 12.2. The protostellar distribution on average looks shifted towards higher values compared to the pre-stellar confirming what we observe in the maps, also in individual clumps (see Appendix~\ref{app:additioal_plots}). This result is confirmed by a Kolmogorov-Smirnov (KS) test \citep{Massey51}. The test yields an \mbox{$p$-value~$< 2\times 10^{-3}$}, which is lower than the statistical significance level of 5\% usually adopted to reject the hypothesis that the two datasets come from the same continuous distribution (e.g., \citealt{Teegavarapu19}). 

\section{Discussion}\label{sec4:results}
The study of CO-depletion in high-mass star-forming regions has been pursued over the years at different scales both via observations and theoretical studies. For example, the global distribution of \fD~reported in Figures~\ref{fig:fD_A} and~\ref{fig:fD_B} is in agreement with the most recent state-of-the-art three-dimensional numerical simulations presented by \cite{Bovino19}, where the authors have simulated the collapse of turbulent and magnetized isothermal cores, exploring different initial conditions. They reported \fD~values between 50-100 on a scale of 2000~au ($\sim$ the effective radius associated with many of the cores identified in ASHES; \citealt{Sanhueza19}), qualitatively in line with our ALMA data. Notably, the values reported by \citet{Bovino19} have been convolved with an ALMA-like point-spread function, showing a loss in the final \fD~of a factor up to three when compared to the original simulated cubes. This might suggest the presence of compact regions where the chemistry of the CO is dramatically influenced by extreme freeze-out conditions not recoverable with our angular resolution.

On the observational side, however, a rigorous comparison with previous results is challenging, since most of the estimates of \fD, whether derived from single-point spectra (e.g., \citealt{Thomas08, Fontani12, Giannetti14}) or maps (e.g., \citealt{Hernandez11, Pon16, Feng16, Feng20, Sabatini19} and \citealt{Gong21}), are obtained at clump scale angular resolutions. Very few exceptions have been reported for high-mass star-forming regions (see \citealt{Zhang09}, \citealt{Morii21} and \citealt{Rodriguez21}). Within this context our results represent the first core-scale interferometric \fD~maps observed with ALMA for a sample of high-mass clumps.

In the specific case of ASHES, \cite{Morii21} estimated \fD~using additional \CeigO(2-1) ASHES data observed in the 70~$\mu$m dark IRDC G023.477+0.114. This source is not included in this work and in the pilot study published in \cite{Sanhueza19}. G023.477+0.114 has a near kinematic distance of 5.2$\pm$0.5~kpc, which implies a linear-scale resolution of $\sim$5900~au, comparable with those of the data presented in Sect.~\ref{sec:observation}. The authors consider a variation of $X_\mathrm{C^{18}O}^\mathrm{E}$ with the galactocentric distance of the source, and constant values for $\gamma=100$. They report average \fD~values between $\sim$40-300 through the 11 cores at different evolutionary stages identified in G023.477+0.114. Notably, also in this case, \fD~does not decrease going from the prestellar to the protostellar stage, and shows extreme values of \fD$>$100 associated with the most evolved sources, in agreement with our findings.

\cite{Zhang09} conducted 1.3~mm spectral line and continuum observations of two massive molecular clumps harboured in the IRDC G28.34+0.06 (\citealt{Pillai06} and \citealt{Wang08}). The \CeigO(2-1) line was observed with the Submillimeter Array (SMA; \citealt{Ho04}) telescope at a resolution of 1.2\arcsec~and with a final sensitivity of 90 mJy beam$^{-1}$ at the spectral resolution of 1.2 km s$^{-1}$. These angular scale corresponds to $\sim$ 4500~au at the source heliocentric distance of 4.5 kpc \citep{Urquhart18}, similar to the physical scales mapped in our ALMA observations. In each clump, the continuum dust emission at 1.3~mm has revealed multiple cores with typical sizes of $\sim$~5000~au. However, out of these cores only one shows a clear detection with an averaged value of \fD~$\sim$~100 \citep{Zhang09}. The authors assumed $\gamma=100$ and $X_\mathrm{C^{18}O}^\mathrm{E} = 5 \times 10^{-9}$. If we rescale the \fD~found for G28.34+0.06, assuming a galactocentric distance of 4.8 kpc in Equations~(\ref{eq:gamma}) and~(\ref{eq:expected_ab})\footnote{Taking a brightness temperature of 2.5 K, a full width at half maximum (FWHM) of 2.5 km s$^{-1}$, a gas temperature of 30 K as reported by \cite{Zhang09} for the detected \CeigO(2-1) line, and the derived a $\gamma = 72$ and $X_\mathrm{C^{18}O}^\mathrm{E} = 5.7 \times 10^{-7}$}, we obtain a \fD~$\sim$~45, which is in line with the range of values reported in Fig.~\ref{fig:fDhist_tot}. Similarly, \cite{Rodriguez21} observed the \CeigO(2-1) line toward the high-mass protostellar candidate ISOSS J23053+5953 SMM2 with SMA at $\sim$2.5\arcsec~($\sim$10$^4$~au at the distance of 4.3 kpc). They report \fD~$\sim$~20, already considering the variation of $\gamma$ and $X_\mathrm{C^{18}O}^\mathrm{E}$ with galactocentic distance of the source ($\sim$10~kpc \citealt{Bosco19}).

It is worth noting that both the sources of \cite{Zhang09} and \cite{Rodriguez21} host mYSOs as demonstrated by the 24~$\mu$m emission peaks and/or the presence of hot molecular cores with already developed outflows/jets. These high values observed in advanced evolutionary stages provide further evidence that on core scales, the degree of CO-depletion may not be suitable to follow the evolution of a core due to the huge amount of cold molecular gas that can surround a mYSO. These high densities could also increase the efficiency of dust grains coagulation \citep{Galametz19}, implying a larger grain size and, in turn, decreasing the heating and subsequent evaporation of CO from the surface of the dust grains \citep{Iqbal18}.

Across the whole clump-scale, however, the revealed chemical picture changes. We have calculated the average \fD~of each clump shown in Fig.~\ref{fig:fD_A} and~\ref{fig:fD_B}, after convolving the $N$(\HH) and $N$(\CeigO) maps to an angular resolution equal to the effective radius of each clump (see Table~\ref{tab:sample}). 
We associate with this derivation a conservative error of 15\% that accounts for the fluctuations observed for \fD~ over the wide range of densities and temperatures found on the molecular cloud scale \citep{Sabatini19}. We have correlated the average \fD~of each source with the corresponding luminosity-to-mass ratio ($L/M$) of the clumps, a well-known distance-independent evolutionary indicator for both low- and high-mass star-forming regions (e.g., \citealt{Saraceno96, Molinari08} and \citealt{Urquhart22}). The bolometric luminosity, $L$, and clump mass $M$, are taken from \cite{Li20} with associated errors of 50\% for $L$ and 20\% for $M$, respectively \citep{Urquhart18}. These uncertainties have been derived from a statistical relevant sample of $\sim$10$^4$ high-mass clumps identified in the APEX Telescope Large Area Survey of the Galaxy (\citealt{Schuller09}).

\begin{figure}
   \centering
   \includegraphics[width=0.95\hsize]{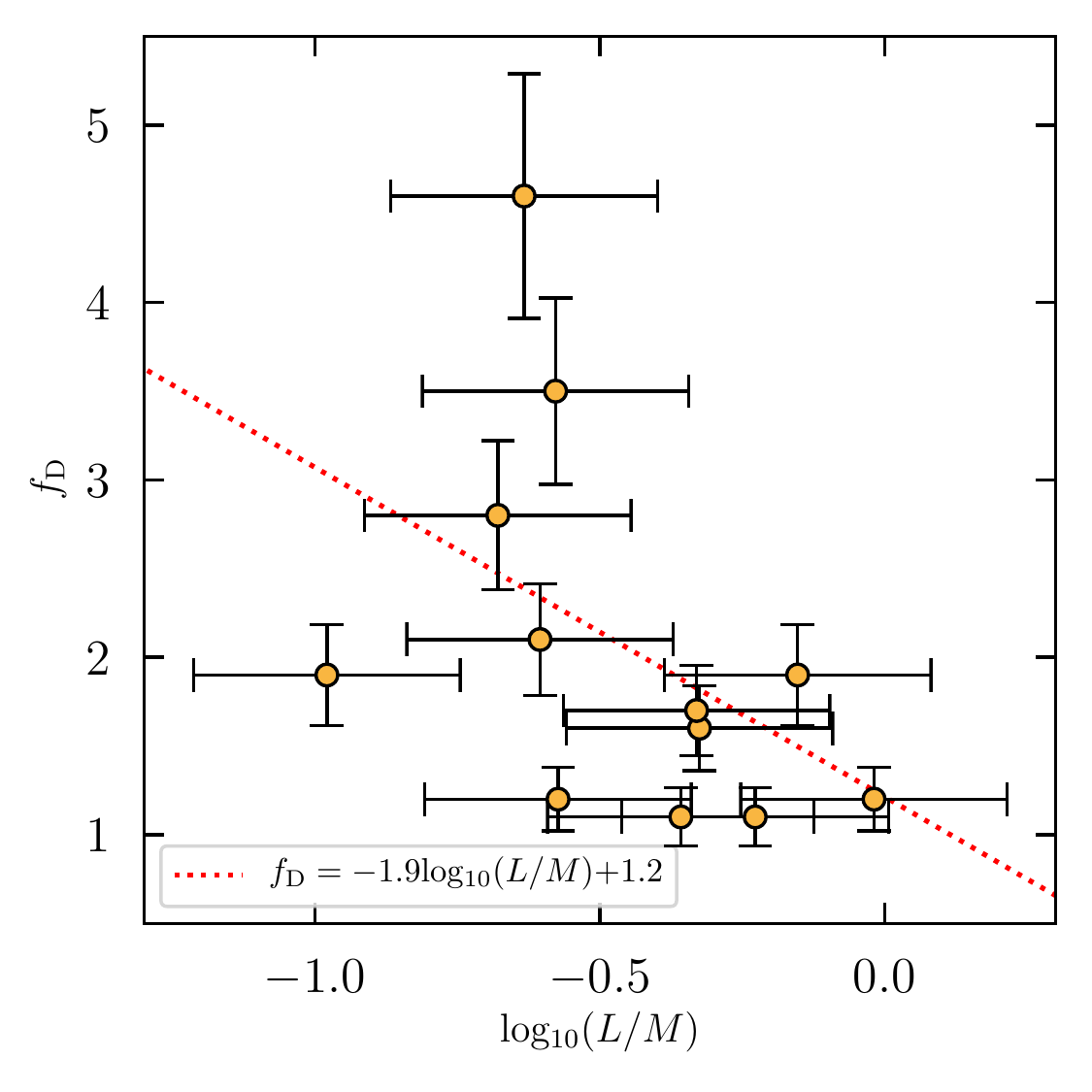}
\caption{Correlation between \fD~and the luminosity-to-mass ratio of the clumps that compose the ASHES sample. Yellow dots are associated with each source, while uncertainties are shown as black bars. Red dashed line represents the linear least-squares fit of \fD~to the log$_{10}$($L/M$). The fit parameters are shown in the legend, with Spearman’s rank correlation coefficient $\rho_{\rm s} =-0.62$ and a $p$-value = 0.03.}\label{fig:fDclumps}%
\end{figure}

Figure~\ref{fig:fDclumps} shows the results of this correlation, presenting a clear downward trend of \fD~with increasing $L/M$. A linear least-squares fit of \fD~to the log$_{10}$($L/M$) is shown as a red-dashed line in the same figure. The averaged \fD~is found to change by a factor of $\sim$5 for about one order of magnitude in $L/M$, yielding the power law relation \mbox{\fD~$= -1.9$log$_{10}$($L/M$)$+1.2$}, with a Spearman’s rank correlation coefficient $\rho_{\rm s} =-0.62$ and a \mbox{$p$-value = 0.03} (e.g., \citealt{Zwillinger00, Cohen13}). This result confirms the reliability of \fD~to classify high-mass clumps at different evolutionary stages, as found in several samples, e.g., \cite{Fontani12, Giannetti14} and \cite{Sabatini19}. The different behaviour of \fD~observed at clump and core-scale, might be the consequence of the complex interplay between chemistry and physics and their associated timescales. In particular, the chemical response to physical changes is smeared out when looking at clump scales, reflecting the average properties of the entire population of cores.

\section{Conclusions}\label{sec5:conclusion}
In this paper we presented the first core-scale \fD~maps derived from \CeigO(2-1) and 1.3~mm continuum ALMA observations for the twelve 70~$\mu$m-dark clumps of the ASHES sample. In this context, we have discussed whether the averaged CO-depletion factor computed at core scales can be considered a reliable evolutionary indicator of the high-mass star formation process.

The overall scenario that emerged from this study shows peculiar chemical conditions for the ISM involved in our targets, that changes according to the physical scale investigated. On the clump scale, we find that on average at least half of the expected CO has been removed from the gas phase, for more than 85\% of the total area mapped in \CeigO. The highest values of CO-depletion are found within the identified cores (both pre- and proto-stellar), where \fD~values of more than 10 are reached in more than $\sim$50\% of the cores.

In contrast to what has been observed for low-mass star-forming cores and, more generally, for high-mass clumps that have the potential to form high-mass stars, our analysis shows that the degree of the CO-depletion process on core-scales does not decrease during the transition from a prestellar to a protostellar phase. If we exclude the temperature effect due to the slight gradients in the $T_{\rm kin}^{\rm NH_3}$ maps, we explain the evolutionary behaviour of \fD~as primarily dependent on the average gas density, which increases with the evolution of the cores. This effect is not observed in the low-mass regime since the high-density regions have smaller sizes and are more affected by temperature variations driven by the star formation process. Furthermore, low-mass star-forming regions are also statistically closer to the Solar System, allowing for a better linear resolution. We emphasise that, due to the poorer resolution of the NH$_3$ maps (on average a factor $\sim$4 coarser) compared to those of \CeigO, temperature is also one of the main uncertainties affecting our results. Our analysis could greatly benefit from ammonia observations with a resolution comparable to that of ALMA. Nevertheless, we highlight the significant improvement for having derived temperatures at $\sim$5\arcsec~angular resolution with respect to adopting the Herschel dust temperatures at 35\arcsec~resolution. 

The \fD~fluctuations appear to be widely distributed when observed over thousands of astronomical units, and in particular trace the densest regions of clumps that are not always associated with a prestellar core. Our results lead us to classify \fD~as a tracer that is not entirely reliable to distinguishing between prestellar and protostellar cores in a high-mass star-forming clumps. However, thanks to the high CO-depletion factors found in large parts of the clump, a more complete picture of evolution can be obtained by observing deuterated molecules. The abundance of ortho-H$_2$D$^+$, for example, has been asserted as a clear chemical indicator of prestellar stages both at the clump scale (e.g., \citealt{Giannetti19, Miettinen20} and \citealt{Sabatini20}) and at the core-scales by the recent results of \cite{Redaelli21} and \cite{RedaelliSUBM} obtained from ALMA data. Improving the sensitivity of astronomical facilities in the mm and sub-mm regime (such as APEX and ALMA) and the systematic study of deuterated molecules (such as H$_2$D$^+$ and D$_2$H$^+$) therefore seem to be the necessary breakthrough to finally obtain a comprehensive picture of the process of high-mass star formation. 

\begin{acknowledgments}
     The authors thank the anonymous Referee, for her/his suggestions to improve the manuscript. GS gratefully acknowledges financial support by the ANID BASAL project FB210003, and Dr R.~Pascale for fruitful discussions. SB is financially supported by ANID Fondecyt Regular (project \#1220033), and the ANID BASAL projects ACE210002 and FB210003. PS was partially supported by a Grant-in-Aid for Scientific Research (KAKENHI Number 18H01259 and 22H01271) of the Japan Society for the Promotion of Science (JSPS). KT was supported by JSPS KAKENHI (Grant Number 20H05645). This paper makes use of the ALMA data ADS/JAO.ALMA\#2015.1.01539.S (PI: P.~Sanhueza). ALMA is a partnership of ESO (representing its member states), NSF (USA) and NINS (Japan), together with NRC (Canada), MOST and ASIAA (Taiwan), and KASI (Republic of Korea), in cooperation with the Republic of Chile. The Joint ALMA Observatory is operated by ESO, AUI/NRAO and NAOJ.
\end{acknowledgments}

\facilities{The Atacama Large Millimeter/submillimeter Array (ALMA; \citealt{Wootten09});}

\software{This research has made use of  \href{http://pyspeckit.bitbucket.org}{{\verb~PySpecKit~}}, \href{http://www.dendrograms.org/}{{\verb~ASTRODENDRO~}} (a~Python package to compute dendrograms of Astronomical data), APLpy (an open-source plotting package for Python; \citealt{Robitaille12}), \href{http://www.astropy.org}{Astropy} (\citealt{Astropy13, Astropy18}), NumPy \citep{Harris20}, Matplotlib \citep{Matplotlib07}, the Cologne Database for Molecular Spectroscopy (CDMS), and the NASA’s Astrophysics Data System Bibliographic Services (ADS);
}

\appendix
\section{C$^{18}$O opacity correction}\label{app:NC18O}
When deriving the $N$(\CeigO), it is worth inspecting whether the emission of \CeigO(2-1) can be assumed optically thin. The optical depth of a transition can be estimated through the peak ratio of the same transition, coming from different isotopologues, if their relative abundance is known (e.g., \citealt{Hofner00}). We computed $\tauCO$ using C$^{17}$O(2-1), C$^{18}$O(2-1) and H$_2$ data published in \citealt{Feng20} observed in G014.492-00.139, and we refer to this paper for a more detailed description of the dataset. We selected this source since the final $N$(\HH) are among the highest of the entire ASHES sample, and therefore G014.492-00.139 represents a ideal case to study the variation of $\tauCO$.

Assuming equal excitation temperatures and filling factor for the two species, $\tauCO$ is estimated as
\begin{equation}\label{eq:ratio}
R_{18,17}=\frac{T^{\rm C^{18}O}_{\rm mb}}{T^{\rm C^{17}O}_{\rm mb}} \propto \frac{[1- {\rm exp}(-\tau_{\rm C^{18}O})]}{[1- {\rm exp}(-\tau_{\rm C^{17}O})]},
\end{equation}
where, $\tau_{\rm C^{17}O}$ is the optical depth of \CsevO(2-1) at \mbox{$\sim$224.7 GHz}, for which we assume $\tau_{\rm C^{17}O}=\tau_{\rm C^{18}O}$/4.16 (e.g., \citealt{Wouterloot08}). The fit is performed pixel-by-pixel for both C$^{17}$O and C$^{18}$O datacubes by taking data where both the continuum and the line emission have a $>$3$\sigma$ level detection. We employ the Python Spectroscopic Toolkit ({\verb~PySpecKit~}; \citealt{Ginsburg11a, Ginsburg22}), by using a single Gaussian component over a velocity space of $\pm$3 km~s$^{-1}$ around the local standard of rest velocities ($V_{\rm lsr}$) derived in \cite{Sanhueza19}. The estimated $\tauCO$ are in the range of $\sim$0.25-1.80, implying optical depth correction factors $C_\tau \in \{1.13-2.16 \}$ to derive the final $N$(\CeigO) from Equation~(\ref{eq:nc18o}). In more than 75\% of the sources detected in \CeigO, $\tauCO < 1.37$ ($C_\tau < 1.84$). Similar $\tauCO$~are also reported in other Infrared Dark Cloud (e.g., \citealt{Sanhueza10, Sabatini19, Gong21}), proving that \CeigO~is virtually always optically thin under the typical conditions prevalent in IRDCs.

To account for the same correction in the other ASHES sources, we follow the same approach of \cite{Sabatini19}, looking for a linear relation between log$_{10}$($\tauCO$) and log$_{10}$[$N$(\HH)]. We have preferred the \HH~column density  over $N$(\CeigO) since $N$(\HH) is not affected by opacity at the observed size scales (i.e., $\kappa_\nu$ correction already applied in Sect.~\ref{sub:NHH}). The final \CeigO~column density maps are derived applying in each source the bestfit \mbox{log$_{10}(\tauCO)$-${\rm log}_{10}[N{\rm (H_2)}]$} relation obtained in G014.492-00.139, i.e., log$_{10}$($\tau_{\rm C^{18}O}$) $=$ 0.6log$_{10}$[$N$(\HH)] --14.4.

\begin{figure*}
   \centering
   \includegraphics[width=1.0\hsize]{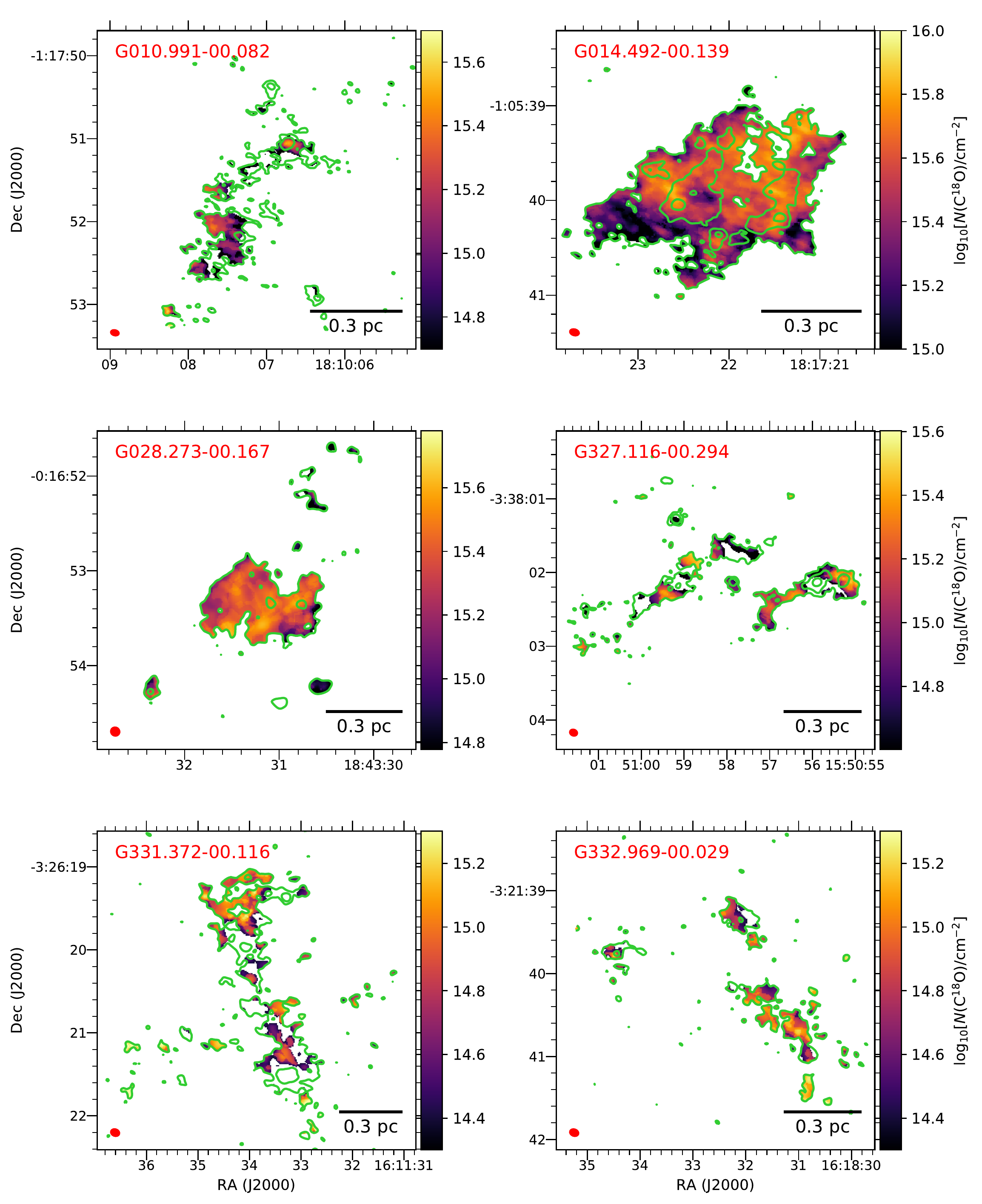}
\caption{Final $N$(\CeigO)~maps obtained following the procedure explained in Sect.~\ref{sub:NCO}. All the maps are corrected for opacity effects as reported in Appendix~\ref{app:NC18O}. Green contours correspond to the ALMA dust continuum emission at [3,9,27]~$\times$~$\sigma$ \citep{Sanhueza19}. The ALMA synthesized beams are displayed in red at the bottom left in each panel, while the scalebar is shown in the bottom right corners. The color wedge of each panel displays the color scales corresponding to $N$(\CeigO)~in the log-scale.}\label{fig:NC18O_A}%
\end{figure*}

\begin{figure*}
   \centering
   \includegraphics[width=1.0\hsize]{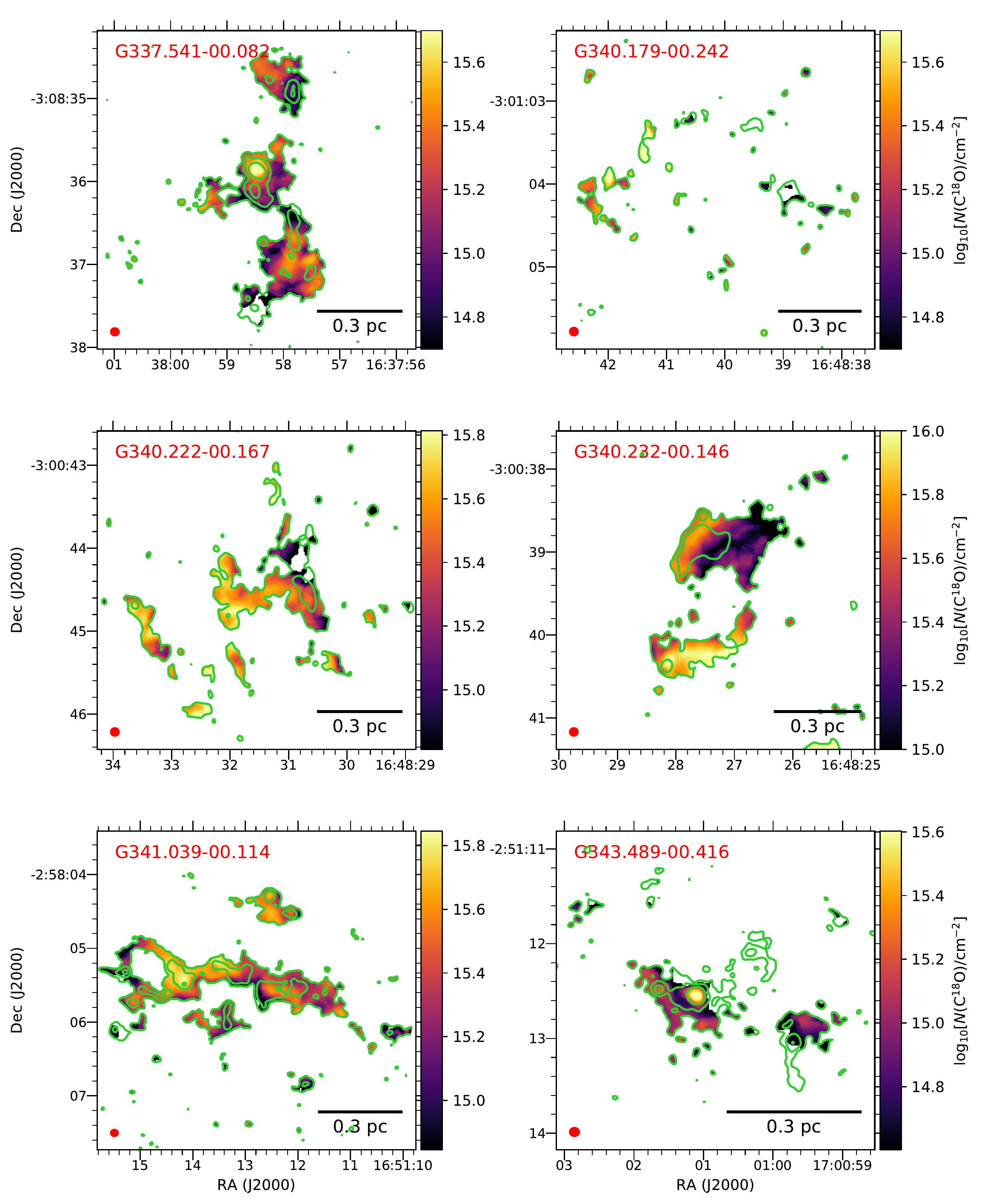}
\caption{Continuation of Fig.~\ref{fig:NC18O_A}.}\label{fig:NC18O_B}%
\end{figure*}

\section{Notes on the analysis for individual clumps}\label{app:additioal_plots}
The aim of this section is to test the influence of some specific properties of the clumps on the results discussed in the sections~\ref{sub:fD_derivation} and~\ref{sec4:results} (e.g., the heliocentric distance of the clumps, the number of cores, or the proportion of pre- and proto-stellar cores found in each ASHES source).
 
\begin{figure}
   \centering
   \includegraphics[width=0.5\hsize]{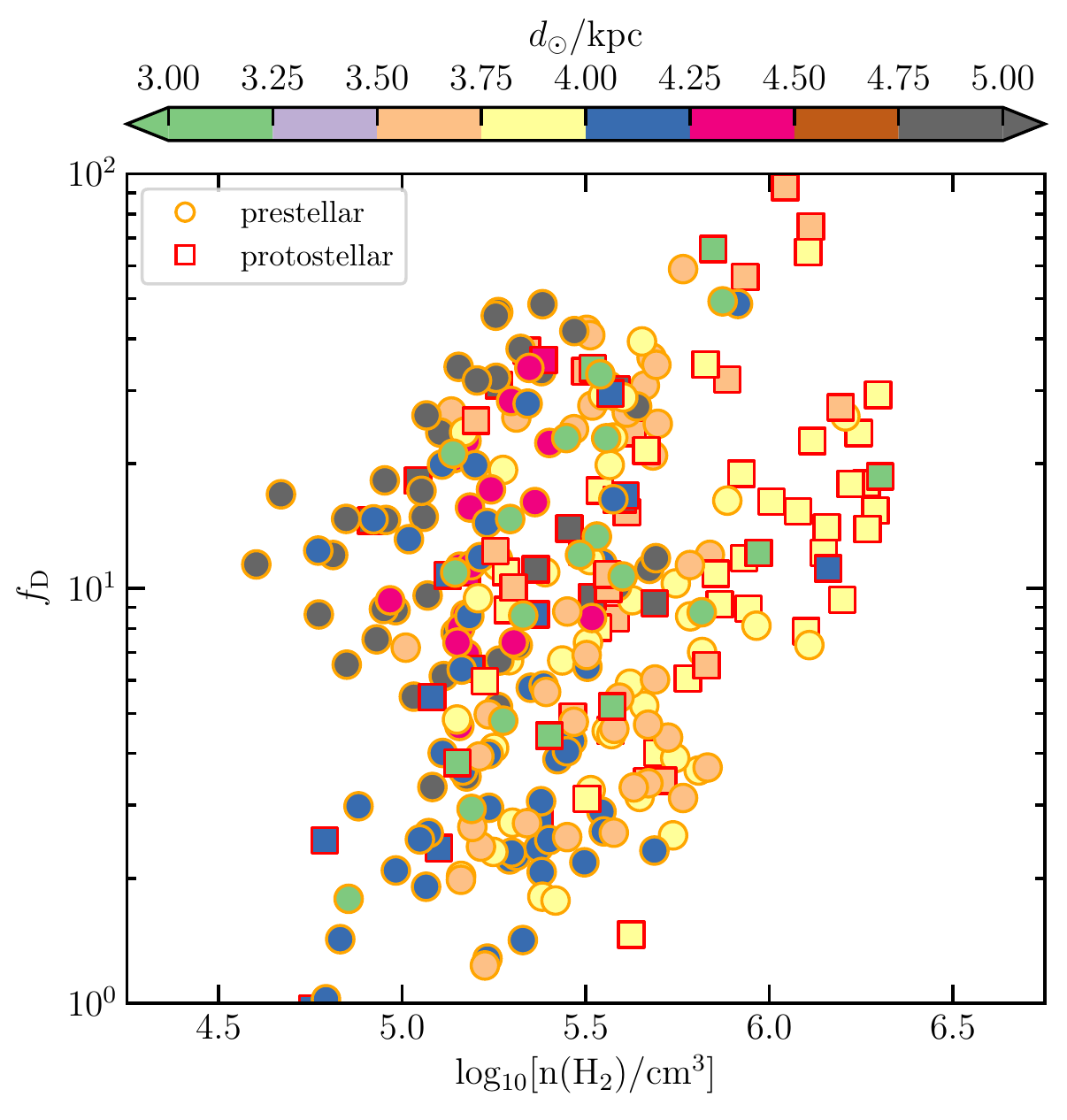}
\caption{Same as Figure~\ref{fig:scatter}($a$) showing the variation in the average \fD~and \nHH~for the core population identified in ASHES as a function of the heliocentric distance of each clump (indicated in the colour wedge; see also Tab.~\ref{tab:sample}). Circles and squares represent the prestellar and protostellar cores, respectively.}\label{fig:scatter_dist}%
\end{figure}

Figure~\ref{fig:scatter_dist} represents the analogue of Fig.~\ref{fig:scatter}, in which we have coloured the cores as a function of the heliocentric distance of the clumps hosting them. There is no clustering of cores when the distance of each source is considered, and the cores associated with each distance bin span comparable ranges of values in terms of \fD~ and \nHH, corresponding to at least a factor of $\sim$5. Thus, the distribution of cores shown in Fig.~\ref{fig:scatter} appears to be distance-independent, ruling out the influence of a possible distance bias on our results.
 
\begin{figure*}
   \centering
   \includegraphics[width=0.9\hsize]{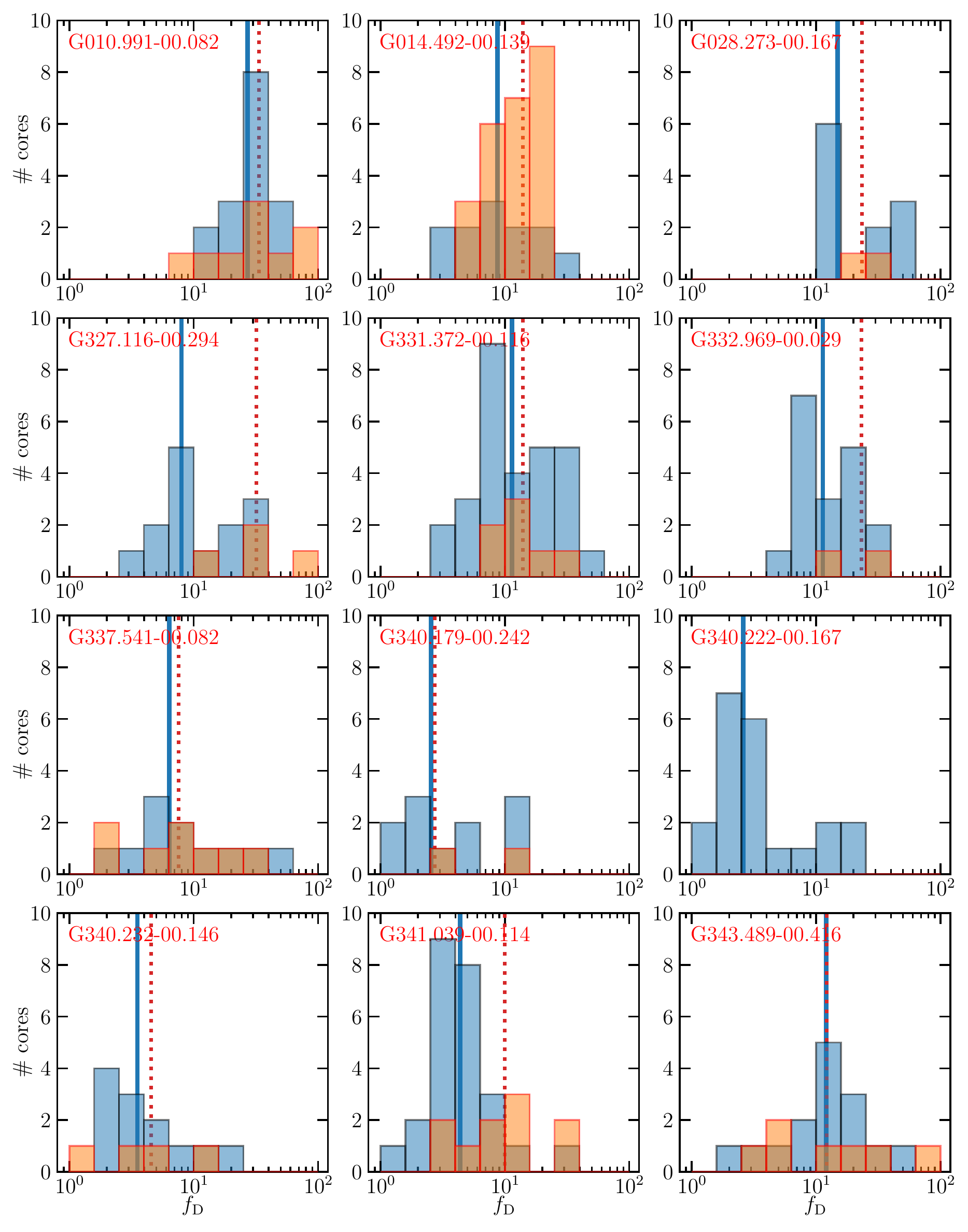}
\caption{Same as Figure~\ref{fig:hist} showing the number distributions of the averaged \fD~ associated with each core identified in ASHES. In the different panels, the distributions within each clump are shown separately (red labels in the upper left corners). The blue and orange histograms refer to prestellar and protostellar cores, respectively. The vertical lines represent the median of \fD~resulting from the distributions of prestellar (blue lines) and protostellar (red-dashed lines) cores.}\label{fig:hist_per_clumps}%
\end{figure*}

Figure~\ref{fig:hist_per_clumps} represents the analogue of Fig.~\ref{fig:hist}, and shows the number distributions of the averaged \fD~associated with ASHES cores within each clump, separately. Although the statistics of the cores in each clump is greatly reduced compared to the total number of cores shown in Fig.~\ref{fig:hist}, we note that the median value of \fD~derived for the prestellar population of cores (blue vertical lines in Fig.~\ref{fig:hist_per_clumps}) is always lower than -- or at most equal to -- the value found for the protostellar cores (red dashed vertical lines; Figure~\ref{fig:hist_per_clumps}). Figure~\ref{fig:hist_per_clumps} also shows that randomly removing a clump from the analysis presented in Sections~\ref{sub:fD_derivation} and~\ref{sec4:results} does not qualitatively change the general conclusions summarised in Sect.~\ref{sec5:conclusion}.

\newpage

\bibliography{mybib_GAL}{}
\bibliographystyle{aasjournal}

\end{document}